\documentclass[aps,prl,reprint]{revtex4-1}

\usepackage{xcolor}
\usepackage{graphicx}
\usepackage{dcolumn}
\usepackage{bm}
\usepackage{ifpdf}
\usepackage{epstopdf}
\usepackage{slashed}
\usepackage{amsfonts}
\usepackage{mathrsfs}
\usepackage{amssymb}
\usepackage{multirow}
\usepackage{CJK}
\usepackage{xcolor}
\usepackage{subfigure,dcolumn}
\usepackage[T2A,T1]{fontenc}
\usepackage[russian,english]{babel}
\usepackage{amsmath}
\usepackage{listings}
\usepackage{booktabs}
\usepackage{color}
\begin{document}

\preprint{APS/123-QED}

\title{Role of the isospin diffusion on cluster transfer in $^{12,14}$C + $^{209}$Bi reactions}

\author{Zepeng Gao$^{1}$}
\author{Yinu Zhang$^{1}$}
\author{Long Zhu$^{1,2,}$}
\email{Corresponding author: zhulong@mail.sysu.edu.cn}
\author{Zehong Liao$^{1}$}
\author{Yu Yang$^{1}$}
\author{Chenchen Guo$^{1}$}
\author{Jun Su$^{1,2}$}

\affiliation{
$^{1}$Sino-French Institute of Nuclear Engineering and Technology, Sun Yat-sen University, Zhuhai 519082, China\\
$^{2}$Guangxi Key Laboratory of Nuclear Physics and Nuclear Technology, Guangxi Normal University, Guilin 541004, China\\
}%

\date{\today}

\begin{abstract}
 Heavy-ion collisions at near-barrier energies provide a crucial pathway for investigating nucleon correlations and clustering structures. 
 Recent experimental results showed that the valence neutrons in light projectiles obviously enhance the $\alpha$ transfer. This finding is extremely puzzled and fascinating, because it violates the ground-state $Q$ value systematics unexpectedly. In this work, the time-dependent Hartree-Fock approach is utilized to investigate the cluster transfer. By comparing the reactions $^{12,14}$C + $^{209}$Bi, we discover that above puzzling behavior is because of the strong correlation between isospin diffusion and clustering. Our calculations clearly show that the equilibrium of neutron-to-proton ratio strongly inhibits the clustering. This work opens a prospect for investigating the clustering in open quantum system. 
 \end{abstract}

\maketitle


\textit{Introduction.} The phenomenon of cluster within atomic nuclei is widely recognized. Clustering has been regarded as a transitional phenomenon between crystalline and quantum-liquid phases of fermionic systems \cite{ebran2012atomic}, which is more prevalent in light nuclei \cite{von2006nuclear,freer2018microscopic}. The strong correlation among nucleons is proposed to play a significant role in maintaining the cluster configuration \cite{carlson1998structure,wiringa2014nucleon}. Especially in the case of symmetric even-even nuclei, like $^8$Be, $^{12}$C, and others, these nuclei can be viewed as being composed of multiple $\alpha$-particles. The second 0$^+$ state of $^{12}$C called Hoyle state, interestingly, which  holds great significance in explaining the abundance of carbon in the universe and understanding the origins of life, dominants degrees of freedom are those of $\alpha$-particle clusters rather than nucleons \cite{freer2014hoyle,otsuka2022alpha,epelbaum2012structure}. Furthermore, extreme spin or isospin can enhance the stability of cluster configurations \cite{zhao2015rod,suhara2010cluster}. High isospin carbon nuclei with cluster configurations of linear-chain molecular band, such as $^{14}$C \cite{han2023nuclear} and $^{16}$C \cite{liu2020positive,han2022observation}, have also been observed experimentally.

Presently, there is growing emphasis on studying the influence of various structural properties of the nucleus during nuclear reactions, including deformed nuclei, neutron skins, halo nuclei, as well as $\alpha$ clusters. Additionally, valuable information about the initial structure can be extracted by analyzing the observables of the reaction's final state. Examples for certain cluster nuclei have been shown that the presence of potential $\alpha$-particles on the surface of Sn isotopes \cite{tanaka2021formation}, as well as the existence of potential four-neutron resonance \cite{duer2022observation}, can be investigated through proton knockout reactions. Furthermore, the clustering phenomenon within atomic nuclei becomes evident in both nuclear giant resonances \cite{he2014giant,he2016dipole} and heavy-ion reaction processes \cite{piantelli2023characterization,frosin2023examination,broniowski2014signatures,zhang2017nuclear,guo2019influence,li2020signatures,dasgupta2023production}. 

The cluster effects become more pronounced at low-energy and are primarily evident in transfer reactions \cite{hodgson2003cluster}. In the case of weakly bound nuclei such as $^6$Li, $^7$Li, and $^8$Be, the ground-state wave functions often exhibit a substantial component comprising the maximum number of particles coupled with the remaining nucleons. Large cross-sections for cluster transfer reactions is a consequence of internal clustering phenomena and low separation energies \cite{canto2009dynamic,canto2015recent,canto2006fusion}. Moreover, in the case of strongly bound nuclei like $^{12}$C and $^{16}$O, a certain degree of cluster transfer cross-sections still remains evident \cite{le1972nuclear,hamada2011analysis,becchetti198016o,jin1980product}. 
In Ref. \cite{sahu2001evidence,biswas2012projectile}, it was observed experimentally that the cross sections of two neutrons correlated cluster transfer are strongly enhanced in the $^{18}$O induced reaction comparing to $^{16}$O with $^{174}$Yb target as well as $^{208}$Pb target. This behavior is extremely puzzled and fascinating, because it violates the ground state $Q$ value systematics unexpectedly. Furthermore, interestingly such enhancement disappear for colliding with the $^{63}$Cu target \cite{shorto2008effects,crema2018reaction}. 
Collisions of the heavy-ions involve a complicated interplay and exchange of energy between relative motion and intrinsic states of nuclei, and the equilibrium of the collective degree of freedom.
In multinucleon transfer reactions, the equilibrium of neutron-to-proton ratio plays an important role on the nucleons rearrangement, which is related to the mechanism of quantum fluctuation and thermal dynamics. The key question that arises is: do the equilibration processes affect the clustering in open quantum systems?

This work is mainly focused on the $\alpha$ and $^{8}$Be clustering. The reactions $^{12,14}$C+$^{209}$Bi reactions are studied. The dynamic process of cluster transfer and the role of the isospin diffusion of fermionic systems on the cluster transfer probabilities are revealed within the framework of the time-dependent Hartree-Fock (TDHF) method.
The TDHF method is a fully microscopic density functional approach \cite{schuetrumpf2018tdhf} that relies on effective nucleon-nucleon interactions, which has found extensive applications in low-energy heavy-ion reactions and fission dynamics \cite{simenel2012nuclear,simenel2018heavy,stevenson2019low,sekizawa2019tdhf}. Furthermore, the structural characteristics of atomic nuclei and their dynamics during nuclear reactions can be uniformly and self-consistently described by only one set of Skyrme parameters \cite{sun2023microscopic,simenel2013microscopic}. Our research focuses on cluster transfer in heavy-ion reactions within the near-barrier energy region, in which nuclear reactions are highly sensitive to the initial structural information and TDHF has the capability to leverage its advantages.

\textit{Theoretical framework}. In this work, the calculation of static nuclei and the dynamic process are described coherently by static HF and TDHF methods, where the clustering effect during the nuclear reaction can be explained self-consistently. The Skyrme density function SLy5 \cite{chabanat1998skyrme} has been utilized in both static and dynamic processes, which has been adopted in many investigations. The static HF are employed using the damped gradient iteration method, and the box size was established to be 24 × 24 × 24 fm$^3$ with a mesh spacing of 1.0 fm, while 50 × 24 × 50 fm$^3$ of box size for dynamical simulation was further fixed. 

\begin{figure}[h!]
\includegraphics[width=0.4\textwidth]{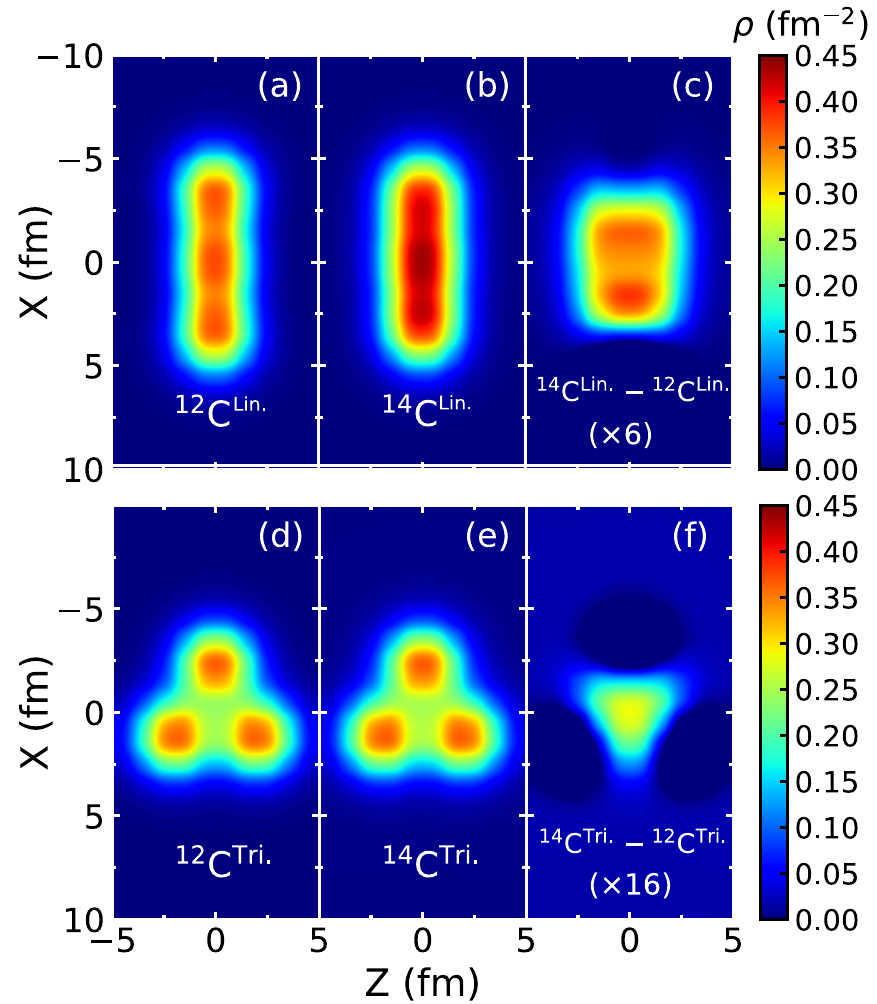}
\caption{\label{fig:1} Two-dimensional density distribution for the linear-chain structure of $^{12}$C$^{\rm Lin.}$(a), $^{14}$C$^{\rm Lin.}$(b) and the difference between them (c) in the static HF calculations and the triangular structure of $^{12}$C$^{\rm Tri.}$(d), $^{14}$C$^{\rm Tri.}$(e) and the difference between them (f) in the DC-HF calculations with the SLy5 parametrization of the Skyrme EDF.}
\end{figure}

A detailed explanation, highlighting its distinction from regular initialization, will be provided for a static process that incorporates the cluster effect. The trial wave function is expressed in the form of Gaussian functions, and two of the 2n-2p (pre-cluster) wave functions are translated upwards and downwards along the x-axis, respectively. The solution with 3$\alpha$ linear-chain configuration for $^{12}$C can thus yield from static HF, where a local minimum energy exists. The two-dimensional density distribution are shown in Fig. \ref{fig:1}(a), where the 3$\alpha$ configuration is shown clearly. Moreover, an equilateral triangular configuration can be obtained using density-constrained HF(DC-HF) \cite{umar2023cluster}, which is shown in Fig. \ref{fig:1}(d). We would like to emphasize that in this work we aim to reveal the effects of N/Z equilibration on the cluster transfer by investigating and focusing on the distinction between $^{12}$C and $^{14}$C involved reactions, even though the linear-chain and triangle are not ground state configurations. The truth is that several studies \cite{otsuka2022alpha, shen2023emergent} suggest that the cluster state of $^{12}$C, adopting a configuration resembling a blunt-angled triangle, may exist in the ground state with a certain probability, allowing the linear-chain and equilateral triangular configurations to be approximated as such. Also, during the collision the dissipated kinetic energy could excite the projectiles, then enhance the probabilities of the molecular states of clusters.

The isospin effects on cluster transfer are investigated by comparing $^{12}$C and $^{14}$C projectiles. The density distribution of $^{14}$C$^{\rm Lin.}$ obtained from static HF calculation is shown in Fig. \ref{fig:1}(b) and $^{14}$C$^{\rm Tri.}$ obtained from DC-HF calculation is shown in Fig. \ref{fig:1}(e). The superscripts "Lin." and "Tri." denote the linear-chain and triangle configurations, respectively. Higher density compared to $^{12}$C$^{\rm Lin.}$ and $^{12}$C$^{\rm Tri.}$, as shown in Fig. \ref{fig:1}(c)(f), is attributed to the contribution of two valence neutrons. Fascinatingly, for valence neutrons in linear-chain structure, after iterating through static HF calculation to reach the local minimum density, the density distribution of these two valence neutrons becomes more concentrated on one side, specifically between the two $\alpha$ particles, even if they were positioned on opposite sides at the beginning. The existence of lower energy and longer linear-chain lifetime in such configuration, as an answer, has been extensively confirmed \cite{ren2020dynamics}. 

For dynamic process, the TDHF equation is written as a set of nonlinear Schr\"{o}dinger-like equations for the occupied single-particle wave functions:
\begin{equation}
i \hbar \frac{\partial}{\partial t} \psi_i(\boldsymbol{r}, t)=h \psi_i(\boldsymbol{r}, t) \quad(i=1, \ldots, A),
\end{equation}
where $h$ is the single-particle Hamiltonian and $A$ is the number of single-particle states. These non-linear equations are solved on a three-dimensional Cartesian grid, allowing for a comprehensive analysis without imposing any symmetry restrictions.

To enhance the analysis of cluster transfer processes, we utilized the density-constrained TDHF (DC-TDHF) method \cite{umar2006heavy} to calculate the interaction potential between the projectile and target nuclei throughout the reaction, which is widely employed in heavy-ion fusion reactions  \cite{sun2022microscopic,godbey2019influence,umar2009density,keser2012microscopic,umar2014energy,godbey2022theoretical,washiyama2015microscopic,jiang2013dynamics,jiang2014microscopic}. At certain times, the instantaneous density is utilized to carry out a static Hartree-Fock (HF) minimization. In this process, the neutron and proton densities are constrained to match the corresponding instantaneous TDHF densities,
\begin{equation}
\begin{aligned}
E_{\rm {D C}}(t)= & \min _\rho\left\{E\left[\rho_n, \rho_p\right]
+\int \mathrm{d} \mathbf{r} \lambda_n(\mathbf{r})\left[\rho_n(\mathbf{r})-\rho_n^{\mathrm{TDHF}}(\mathbf{r}, t)\right]\right. \\
& \left.+\int \mathrm{d} \mathbf{r} \lambda_p(\mathbf{r})\left[\rho_p(\mathbf{r})-\rho_p^{\mathrm{TDHF}}(\mathbf{r}, t)\right]\right\} .
\end{aligned}
\end{equation}
One can thus define the ion–ion potential as:
\begin{equation}
V(R)=E_{\rm {D C}}(R)-E_1-E_2,
\end{equation}
where $E_1$ and $E_2$ are energies of projectile and target nucleus obtained from static HF calculations, respectively. The excitation energy in nuclear reactions can be expressed as:
\begin{equation}
E^*(R)=E_{\rm TDHF}(R) - E_{\rm DC}(R) - V(R).
\end{equation}

Furthermore, the particle number projection (PNP) method \cite{simenel2010particle}, which is widely applied in multinucleon transfer reactions \cite{evers2011cluster,sekizawa2013time,sekizawa2016time,sekizawa2017microscopic,wu2022production,wu2019microscopic,jiang2020probing}, will be employed to calculate the probability of cluster transfer. The transfer probabilities in transfer reactions can be obtained from the post-collision TDHF wave function, in a spatial volume $V$ containing the fragment, onto $N$ protons or neutrons that represent a certain transfer channel. The projection operator can be expressed as an integral over the gauge angle $\theta$, in the following form:
\begin{equation}
\hat{P}_V^q(N)=\frac{1}{2 \pi} \int_0^{2 \pi} {\rm e}^{i \theta\left(\hat{N}_V^q-N\right)}{\rm d} \theta.
\end{equation}
The probability of finding $N$ particles with an isospin of $q$ in the volume region $V$ is:
\begin{equation}
P_V^q(N)=\frac{1}{2 \pi} \int_0^{2 \pi}  e^{-i \theta N}\left\langle\Psi\left|{\rm e}^{i \theta \hat{N}_V^q}\right| \Psi\right\rangle {\rm d} \theta.
\end{equation}
The probability of $^4$He and ${^8}$Be transfers can thus be expressed as:
\begin{equation}
\begin{aligned}
   & P(^4{\rm He})=P^{\rm n}(N_0-2)P^{\rm p}(Z_0-2) , \\
   & P(^8{\rm Be})=P^{\rm n}(N_0-4)P^{\rm p}(Z_0-4) ,
\end{aligned}
\end{equation}
where $N_0$ and $Z_0$ denote neutron number and proton number of $^{12}$C or $^{14}$C, respectively.

\begin{figure}[h!]
\includegraphics[width=0.5\textwidth]{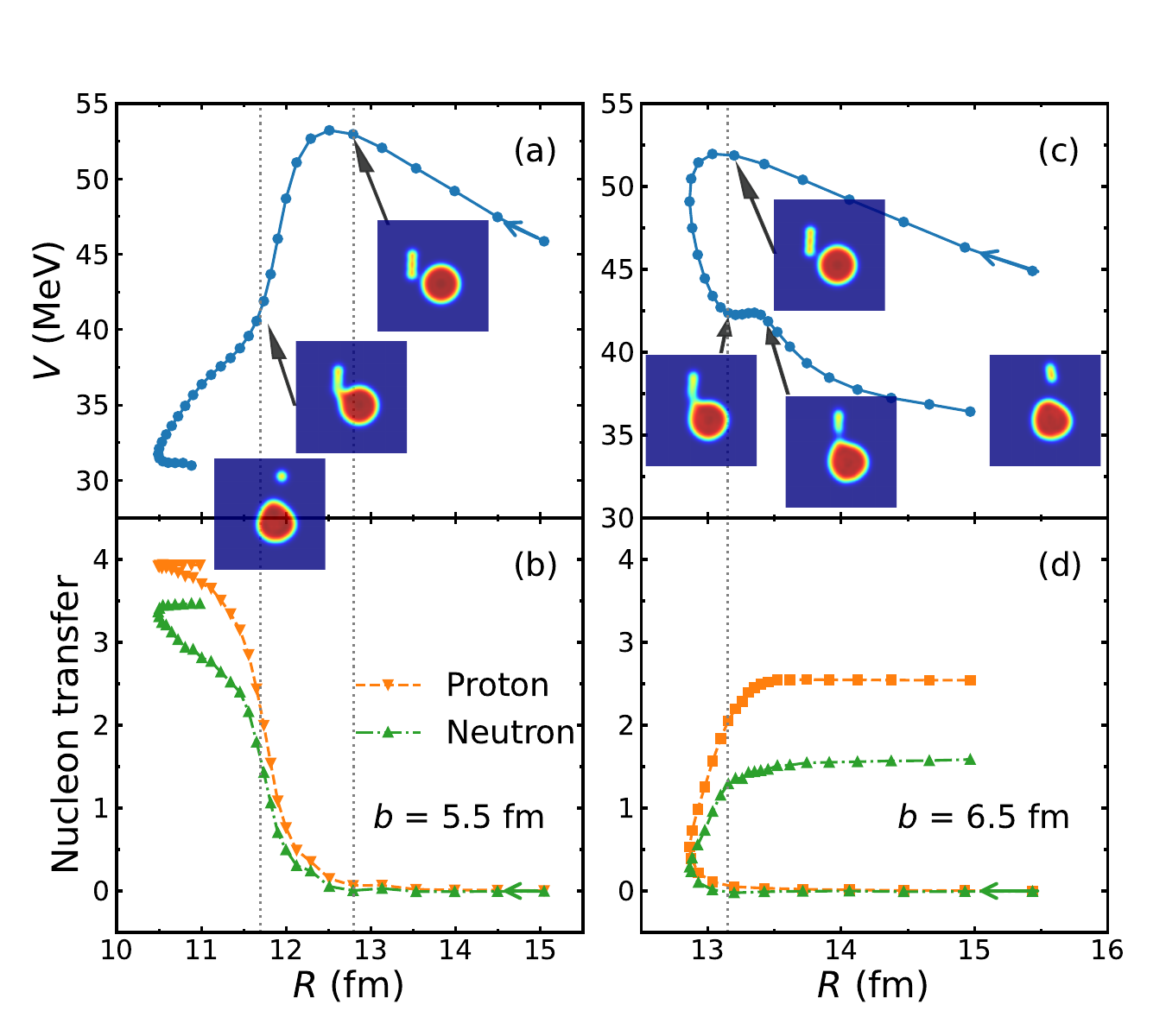}
\caption{\label{fig:2} The DC-TDHF potential (upper panels) calculated for the $^{12}$C$^{\rm Lin.}$+$^{209}$Bi at $E_{\rm c.m.} = 75$ MeV and the corresponding particle number (lower panels) of nucleon transfer from of projectile to target as functions of distance of the center-of-mass at impact parameter $b = 5.5$ fm (a),(b) and $b = 6.5$ fm (c),(d). The snapshots of density distribution are also shown.}
\end{figure}

\textit{Results and discussions.} The interaction potential and nucleon transfer in the reaction of $^{12}$C$^{\rm Lin.}$+$^{209}$Bi at $E_{\rm c.m.} = 75$ MeV are shown in  Fig. \ref{fig:2}. With an impact parameter of $b = 5.5$ fm, the ion-ion interaction potential initially increases as the center-of-mass distance decreases. At $R=12.8$ fm, near the range of nuclear forces, the potential energy starts to decrease. This signifies the initiation of mass transfer between the two nuclei as shown in Fig. \ref{fig:2} (b). When $R=10.5$ fm, the projectile-like fragment departs from the target with a significant portion of mass transferring from $^{12}$C$^{\rm Lin.}$ to $^{209}$Bi, while approximately one $\alpha$ particle scattered. Notably, there are two turning points($R=11.7$ fm and $R=10.5$ fm) in the descending potential curve, precisely corresponding to the transfer of two and four protons from $^{12}$C$^{\rm Lin.}$ to the target (see Fig. \ref{fig:2} (b)). Thus, this turning point provides compelling evidence for cluster transfer.

For the impact parameter of 6.5 fm, a turning point is also observed. Due to the grazing nature of the collision, the duration of the nuclear reaction is reduced, leading to a relatively small number of transferred nucleons. Intriguingly, a notable disparity between numbers of the transferred protons and neutrons is shown. In grazing collisions, the substantial neutron skin of $^{209}$Bi results in a significant isospin asymmetry in the neck. This, in turn, enhances the effects of symmetry energy and promotes proton transfer. Therefore, one conjecture can be made that this behavior arises from the competition between cluster transfer and isospin diffusion.

\begin{figure}[h!]
\includegraphics[width=0.4\textwidth]{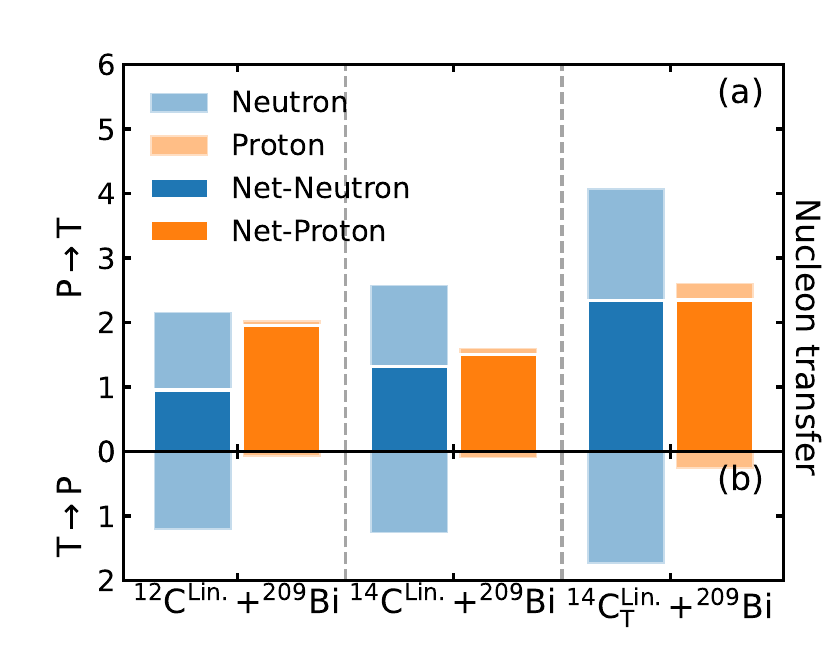}
\caption{\label{fig:3} The nucleon transfer number as functions of three reaction systems at $E_{\rm c.m.}$ = 70 MeV, $b$ = 6 fm. The nucleon transfer from the projectile to the target (a), from the target to the projectile (b) and the net transfer number are both shown.}
\end{figure}

The isospin diffusion effects in the reaction process can be explained by the bidirectionality of nucleon transfer. To clarify the role of isospin diffusion on the cluster transferring, ${^{14}}$C is compared to ${^{12}}$C in the collisions with $^{209}$Bi. The presence of two valence neutrons in ${^{14}}$C outside the ${^{12}}$C core is expected to influence the nucleon transfer process. The number and the direction of nucleon transfer which obtained from the integration for specific wave functions are illustrated in Fig. \ref{fig:3}. Note that there is a mere 0.5 MeV difference in the Coulomb barriers of the two systems. As a result, we employ identical incident energies to contrast the impacts related to isospin and valence neutrons. For ${^{12}}$C$^{\rm Lin.}$+$^{209}$Bi, about two protons and two neutrons are transferred from the projectile to the target, suggesting that microscopic analysis of transfer reactions can involve considering them as the concurrent transfer of both a proton pair and a neutron pair \cite{rotter1969reactions,kurath1974alpha}. Meanwhile, a significant neutron reverse-transfer (from $^{209}$Bi to ${^{12}}$C$^{\rm Lin.}$) takes place, while there is almost no proton flow from $^{209}$Bi to ${^{12}}$C$^{\rm Lin.}$. This is because the neck region is neutron-rich due to the neutron skin of $^{209}$Bi and the isospin diffusion plays an important role. 

On the other hand, for $^{14}$C$^{\rm Lin.}$ with the valence neutrons, the net neutron transfer to the target is enhanced. The valence neutrons of 0.7 transfer to the target has been find, which can offset a significant portion of the reverse-transfer neutron. Moreover, the neutron-proton ratio of $^{14}$C$^{\rm Lin.}$ more closely resembles $^{209}$Bi than $^{12}$C$^{\rm Lin.}$, leading to less proton transfer to the target. The combined effects of the aforementioned factors influence the net nucleon transfer, ultimately resulting in one neutron-proton pair of net transfer from $^{14}$C$^{\rm Lin.}$. 

\begin{figure}[h!]
\includegraphics[width=0.45\textwidth]{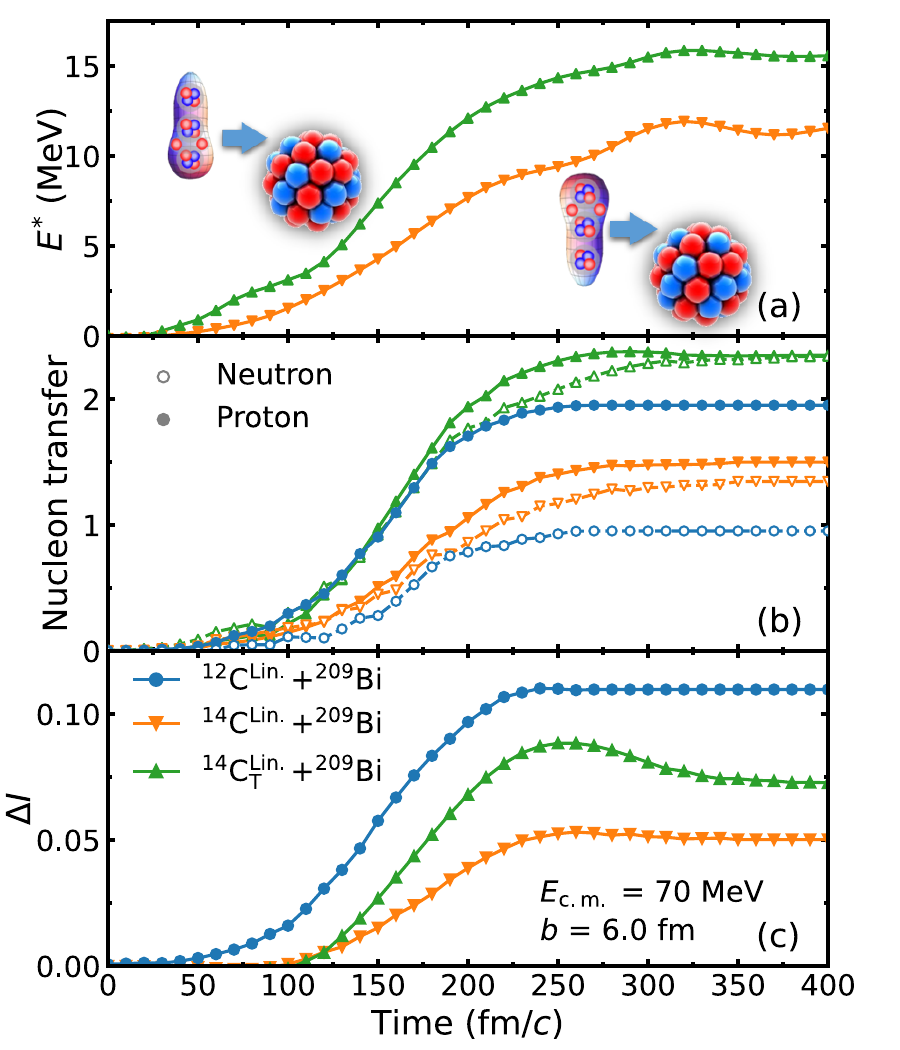}
\caption{\label{fig:4} Time evolution of the excitation energy from DC-TDHF (a), the net nucleon transferring number from the projectile to the target (b) and the isospin variation of projectile-like fragments (c) for $^{12}$C$^{\rm Lin.}$+$^{209}$Bi (circles), $^{14}$C$^{\rm Lin.}$+$^{209}$Bi (lower triangles) and  $^{14}$C$_{\rm T}^{\rm Lin.}$+$^{209}$Bi (upper triangles) at $E_{\rm c.m.}$ = 70 MeV, $b$ = 6 fm.}
\end{figure}

We also show the results in the reaction $^{14}$C$_{\rm T}^{\rm Lin.}$+$^{209}$Bi. Here, $^{14}$C$_{\rm T}^{\rm Lin.}$+$^{209}$Bi denotes the configuration that the two valence neutrons are more concentrated near the neck region during the collision compared to $^{14}$C$^{\rm Lin.}$. The $^{14}$C$_{\rm T}^{\rm Lin.}$ is obtained from performing a spatial inversion along the $x$-axis on the wave function of $^{14}$C$^{\rm Lin.}$. More intense nucleon transfer and exchange are shown. As illustrated in Fig. \ref{fig:4}(a)(b), because of high neutron-richness around the neck region, $^{14}$C$_{\rm T}^{\rm Lin.}$ induced reaction present larger energy dissipation, as well as about one more neutron-proton pair transfer to the target in comparison to $^{14}$C$^{\rm Lin.}$. This means that a roughly alpha particle is purely transferred from $^{14}$C$_{\rm T}^{\rm Lin.}$ to $^{209}$Bi. However, for the reaction $^{12}$C$^{\rm Lin.}$+$^{209}$Bi, the number of pure proton transfer is almost two times more than that of neutron transfer. As illustrated in Fig. \ref{fig:4}(c), for the case of $^{12}$C$^{\rm Lin.}$ the value of $\Delta I$ (where $\Delta I = I-I(t=0)$, $I=(N-Z)/(N+Z)$) increases rapidly compared to those of the $^{14}$C$^{\rm Lin.}$ and $^{14}$C$_{\rm T}^{\rm Lin.}$ configurations. The isospin diffusion weakens the proton and neutron paring during the collisions, which consequently could lower down the probabilities of the alpha cluster transfer.

\begin{figure}[h!]
\includegraphics[width=0.5\textwidth]{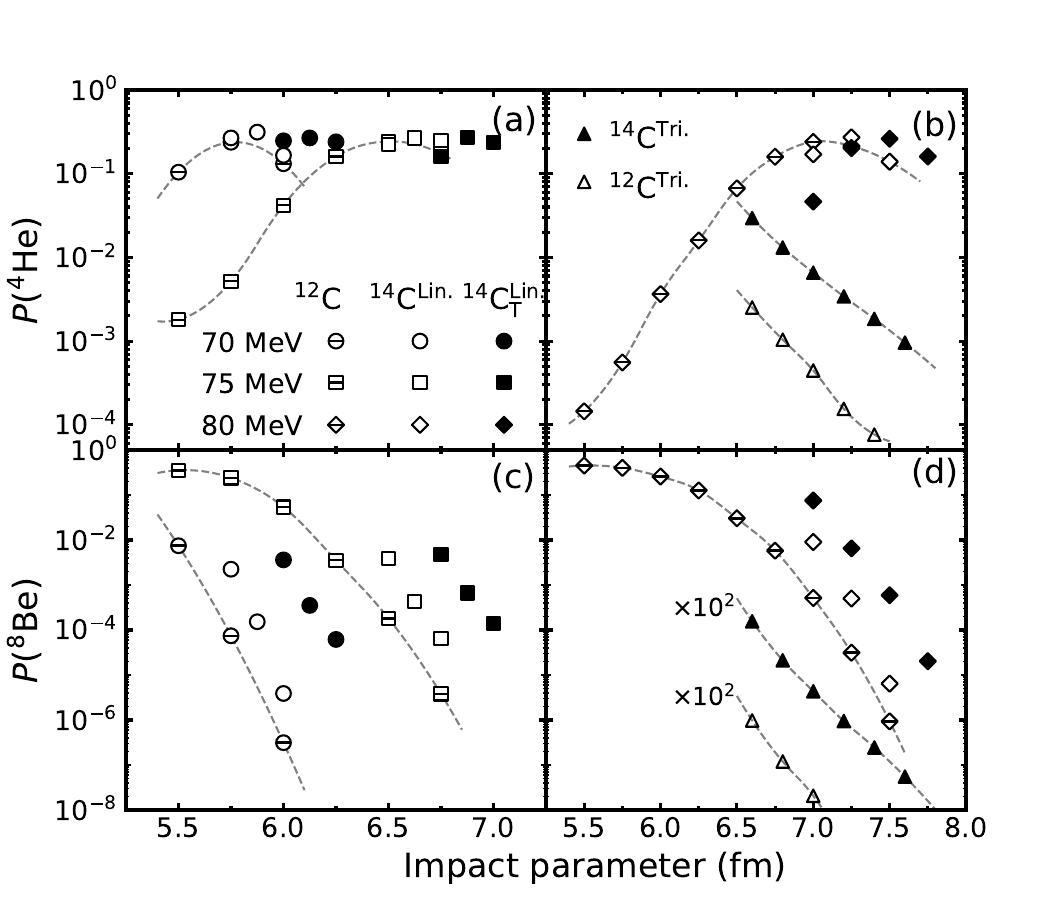}
\caption{\label{fig:5} The transfer probabilities of $^{4}$He (upper panels) and $^{8}$Be (lower panels) as functions of impact parameter $b$. The dots with a horizontal line, open dots and solid dots denote $^{12}$C, $^{14}$C and $^{14}$C$^{\rm T}$ projectiles, respectively. Circles, squares and diamonds denote $E_{\rm c.m.} = 70$ MeV, 75 MeV and 80 MeV, respectively. The open triangles and solid triangles denote $^{12}$C$^{\rm Tri.}$ and $^{14}$C$^{\rm Tri.}$ at $E_{\rm c.m.}$ = 80 MeV, respectively. The lines are only to guide the eyes.}
\end{figure}

To further clarify the effects of isospin diffusion on the cluster transfer, we quantitatively show the probabilities of transferring clusters of $^{4}$He and $^{8}$Be in Fig. \ref{fig:5}. 
Owing to the spatial dispersion characteristic of the wave function, the probabilities of cluster transfer can be extracted by using the PNP method. 
It can be seen that, at same incident energy and impact parameter, $^{14}$C$^{\rm Lin.}$ shows higher cluster transfer probability than $^{12}$C$^{\rm Lin.}$, especially for transferring the $^8$Be cluster, although $^{209}$Bi($^{12}$C,$^{4}$He)$^{217}$Fr having a higher $Q_{\rm gg}$ of $-21.6$ MeV compared to $^{209}$Bi($^{14}$C,$^{6}$He)$^{217}$Fr of $-33.8$ MeV. The cluster transfer in $^{12}$C$^{\rm Lin.}$ induced reaction, as shown in Fig. \ref{fig:3}, is restrained by the reverse-transfer of neutron from the target nucleus, an effect that becomes less pronounced in the case of $^{14}$C$^{\rm Lin.}$ due to the presence of two valence neutrons, especially for the configuration of $^{14}$C$_{\rm T}^{\rm Lin.}$. We would like to emphasize that the Coulomb barriers and elongation along the $x$ direction are very close for $^{12}$C and $^{14}$C induced reactions. Therefore, it is reasonable to compare these reactions at same incident energy and with same impact parameter. To verify that the above behavior does not depend on the configurations, we also show the results in the triangular case. The strong enhancement of $^8$Be transfer in $^{14}$C$^{\rm Tri.}$ induced reaction is clearly shown. Furthermore, the $^4$He transfer probabilities are also much larger in the reaction $^{14}$C$^{\rm Tri.}$ + $^{209}$Bi because of the valence neutrons. The obvious behavior of the $^4$He and the $^8$Be transferring probabilities
testifies our conjecture that the isospin diffusion strongly inhibits the cluster transferring. 


\textit{Summary.} We find the enhancement of the cluster transfer in the reactions involving the light projectile with valence neutrons. The same behavior also had been noticed experimentally. However, why does it violate the ground state $Q$ value systematics? Do the equilibrium processes affect the clustering in open quantum systems? In this work, the fully microscopic approach TDHF was employed to investigate the role of the isospin diffusion on the cluster transfer by comparing the $^{12,14}$C+$^{209}$Bi systems. The evidence of cluster transfer phenomena, by analyzing the nuclear interaction potential and nucleon transfer, was obtained.
A bidirectional neutron transfer process has been observed, exerting a dampening effect on cluster transfer probabilities for the $^{12}$C+$^{209}$Bi system. In contrast, within the projectile of $^{14}$C, the transfer of valence neutrons partially offsets the reverse-transfer of neutrons from the target nucleus. Enhanced energy dissipation, prolonged contact time, and high neutron-richness near neck region results in an elevated cluster transfer probability. The quantitative comparisons of $^{12}$C and $^{14}$C in both the linear-chain and triangular configurations certify the strongly correlation between the isospin diffusion and the clustering. It is found for the first time that the process of isospin diffusion results in a hindrance of cluster transfer. 
This is one direct evidence that the statistical equilibration processes affect the clustering in open quantum many-body systems.

\textit{Acknowledgments.} The authors thank C. J. Lin, H. M. Jia, P. W. Wen, P. W. Zhao and D. D. Zhang for useful discussions. This work was supported by the National Natural Science Foundation of China under Grants No. 12075327; The Open Project of Guangxi Key Laboratory of Nuclear Physics and Nuclear Technology under Grant No. NLK2022-01; Fundamental Research Funds for the Central Universities, Sun Yat-sen University under Grant No. 23lgbj003.
\bibliography{main}

\end{document}